\documentclass{article}
\usepackage[english]{babel}
\usepackage[a4paper,top=2cm,bottom=2cm,left=3cm,right=3cm,marginparwidth=1.75cm]{geometry}
\usepackage{amsmath}
\usepackage{graphicx}
\usepackage{setspace}
\usepackage{authblk}
\usepackage{siunitx}
    \DeclareSIUnit\a{\text{Å}}
    \DeclareSIUnit\angstrom{\text{Å}}
    \DeclareSIUnit\bar{bar}
    \DeclareSIUnit\wt{\%wt}
    \DeclareSIUnit\v{\%v}
\usepackage[T1]{fontenc}
\usepackage{indentfirst}
\usepackage{cite}

\title{
Nanostructure of PEGDA-PEG hydrogel membranes and how it controls their permeability}
\date{}

\author[1,2]{Sixtine de Chateauneuf-Randon}
\author[1,2]{Malak Alaa Eddine}
\author[1]{Bruno Bresson}
\author[3]{Thomas Salez}
\author[6]{Sylvain Prévost}
\author[2]{Sabrina Belbekhouche}
\author[1]{Théo Merland}
\author[4]{Cédric Lorthioir}
\author[2,5]{Clémence Le Coeur}
\author[1]{Cécile Monteux\thanks{Email: cecile.monteux@espci.fr}}

\affil[1]{Laboratoire Sciences et Ingénierie de la Matière Molle, CNRS UMR 7615, PSL Université, Sorbonne Université, ESPCI Paris, 10 rue Vauquelin, Cedex 05 75231 Paris, France}
\affil[2]{Université Paris-Est Creteil, CNRS, Institut Chimie et Matériaux Paris-Est, UMR 7182, 2 Rue Henri Dunant, 94320 Thiais, France}
\affil[3]{University of Bordeaux, CNRS, LOMA, UMR 5798, Talence F-33400,
France}
\affil[4]{Laboratoire de Chimie de la Matière Condensée de Paris, CNRS UMR 7574, Sorbonne Université, 4 place Jussieu, 75005 Paris, France}
\affil[5]{Laboratoire Léon Brillouin, CEA-CNRS (UMR 12), CEA Saclay, Université Paris-Saclay, 91191 Gif-sur-Yvette Cedex, France}
\affil[6]{Large-Scale Structures Group, Institut Laue-
Langevin, 38042 Grenoble, France}

\begin{document}

\maketitle
\newpage
\begin{abstract}

The spacial heterogeneity of hydrogels composed of PEGDA and added polymer chains is expected to play a crucial role on their transport properties which can be exploited in filtration or tissue engineering. However little is known about the arrangement of the polymer chains in the matrix and the length scales of these heterogeneities. Here we combine solid-state NMR and Small Angle Neutron Scattering to unravel the structure and dynamics of PEGDA hydrogels containing added PEG chains of various concentrations. Our results show that the samples present heterogeneities in both the PEGDA and PEG concentrations and suggest that the PEG chains entangle with the PEGDA network. When plotting the sample permeability, $K$, as a function the specific surface of the PEGDA heterogeneities we obtain a master curve, showing that the heterogeneity of the PEGDA matrix controls the permeability of the sample. Moreover the scaling $K \propto V/S$ suggests a structure composed of facetted PEGDA/PEG heterogeneities separated by a network of aqueous thin and flattened films in which the water can permeate.

\end{abstract}

\section{Introduction}

Hydrogels with multiscale porosity have emerged as versatile materials with applications in fields such as bone repair\cite{bai2018bioactive,yue2020hydrogel}, tissue engineering\cite{kumar2017pva,yang2020hydrogel}, and, more recently, filtration\cite{de2018structured,gu2020monolithic}. Controlling the porous structure of hydrogels is crucial to optimize their transport properties such as nutrient diffusion, permeation and cellular growth \cite{li2024emulsion,abune2019macroporous}. 
A large effort has been devoted to develop methods to obtain materials with controlled porosity, from emulsion or particle templating, to polymerization induced phase separation \cite{Foudazi2022, duvsek1965phase, duvsek1967phase}. 
Among macroporous hydrogels designed for filtration applications, those based on poly(ethylene glycol diacrylate) (PEGDA)  hydrogels have gained significant attention in recent years \cite{Ju2009, Dusek, lin2005structure, waters2010morphology, ju2009preparation, Wu2010} especially those in which free PEG chains are added \cite{Sadeghi2018un,decock2018situ,kang2008preparation, cappello2020microfluidic, eddine2022large, eddine2024tuning, eddine2023sieving}.  Indeed such PEGDA hydrogels containing added PEG chains combine good mechanical strength and a permeability which can be tuned by adjusting the free PEG concentration and molar mass \cite{Decock2018,eddine2024tuning}. In our group\cite{eddine2024tuning,eddine2022large,eddine2023sieving}, we have demonstrated that the added PEG chains do not behave as simple porogens that are rinsed out of the membrane after the photopolymerization, leaving PEG-free defects in the network. Instead, these chains remain trapped within the membranes. Our work also shows that \cite{eddine2024tuning,eddine2022large,eddine2023sieving} the addition of PEG chains in a PEGDA matrix controls the membrane's permeability. Upon increasing the PEG concentration, the permeability increases and reaches a maximum for a PEG concentration equal to the overlap concentration ($C^*$)  of the PEG chains in solution. Upon further increase of the PEG concentration the permeability decreases. This raises the question of whether this variation in permeability is linked to structural changes within the membrane. 
Despite numerous studies on PEGDA-based macroporous hydrogels, little is known regarding the spacial arrangement of PEG and PEGDA. In many experiments, PEG is often treated simply as an inert porogen which leaves the matrix after a rinsing step, and researchers mostly rely on indirect observations—like turbidity, permeability, or surface imaging to infer the structure. 

To tackle these questions, we used a combination of isotopic labeling and Small Angle Neutron Scattering to selectively obtain the arrangement of the PEG chains and the PEGDA matrix. We also performed solid-state NMR measurements to get information on chain dynamics. Altogether, these methods allow us to directly observe structural heterogeneities in these PEGDA/PEG hydrogel membranes.

Our data reveal that PEGDA/PEG hydrogels exhibit large-scale compositional fluctuations, and the chains show restricted mobility, likely due to entanglement between the added PEG chains and the PEGDA matrix. These observations suggest that PEG’s role is not limited to serving as a porogen—it actively influences the network structure and dynamics. This should be taken into account when designing hydrogels with controlled transport or mechanical properties.

\section{Materials and Methods}

\subsection{Materials}

The membranes are synthesized from poly(ethylene glycol) diacrylate oligomers (PEGDA, $\overline{\mathrm{M_w}} =\SI{700}{\g\per\mol}$), 4-(2-hydroxyethoxy)phenyl-2-hydroxy-2-propyl ketone (Irgacure 2959) used as a photoinitiator, and linear polyethylene glycol (PEG;$ \overline{\mathrm{M_w}} = \SI{35}{\kilo\g\per\mol}$; dispersity index \DH~$ = 1.1$, see SI Table S1). These products were purchased from Sigma–Aldrich. Water was purified using a Milli-Q reagent system (Millipore). For the small-angle scattering experiments, deuterated PEG ($\alpha,\omega$-bis(deuteroxy)-terminated poly(ethylene glycol-d4), $  \overline{M_w} = \SI{35}{\kilo\g\per\mol}$, \DH~$= 1.2$, Polymer Source\textsuperscript{\texttrademark}, Inc.), as well as deuterium oxide (D$_2$O, \SI{99.9}{\percent} MagniSolv\textsuperscript{\texttrademark}, Sigma–Aldrich) were used. All polymers were analyzed by SEC to obtain the molar masses and the polydispersity index (\DH) indicated above (see Supporting Information).

\subsection{Sample fabrication and composition}

The membranes were prepared using a procedure similar to that described in previous research \cite{eddine2022large, eddine2023sieving, eddine2024tuning}. Briefly, for each formulation, a PEGDA/initiator solution was prepared by dissolving \SI{0.1}{\wt} of Irgacure in 10 mL of PEGDA. This solution was then stirred for 10 minutes. Aqueous solutions composed of \SI{20}{\wt} of the PEGDA/photoinitiator mixture and various contents of PEG were prepared. Examples of PEGDA/PEG mixtures in water are provided in SI, Table S2. The final concentration of the photoinitiator was therefore \SI{0.02}{\wt}. For both NMR and SANS experiments, the PEG chains were either hydrogenated (PEG-h) or deuterated (PEG-d), as explained later. The solvent was H$_2$O, D$_2$O, or a mixture of H$_2$O/D$_2$O in a volumetric ratio of 0.78/0.22. This ratio was chosen to suppress the scattering from the PEGDA matrix in the SANS measurements (See SI Fig. S1 and S2). After overnight stirring, the PEGDA/PEG/initiator solution was poured, bubble-free, between two glass plates separated by a \SI{1}{\milli\m}-thick spacer. The free radical photopolymerization of PEGDA was performed under UV irradiation (Herolab UV - 30L, Intensity = \SI{1800}{\micro\watt\per\centi\m^2}, $\lambda = \SI{365}{\nano\m}$) for 15 minutes. After polymerization, the resulting hydrogel was placed in a Petri dish filled with the used-solvent for at least 24 hours prior to use, to remove any unreacted PEGDA oligomers or untrapped polymer chains. Liquid samples were stirred at least 4 hours prior to the experiments. A solvent exchange from H$_2$O/D$_2$O to H$_2$O was performed for the PEGDA/PEG-d membranes by stirring the polymerized membrane in a large volume of H$_2$O and replacing the supernatant every day for one week. The NMR experiments were carried out on the same samples as the ones considered for the SANS measurements, thanks to this exchange. 

\subsection{Small-Angle Neutron Scattering (SANS)}

Small-Angle Neutron Scattering (SANS) data were recorded using the D22 instrument\cite{websiteD22} (10.5291/ILL-DATA.9-11-2107). Two detectors were employed with a wavelength of $\lambda = \SI{6}{\angstrom}$, covering a $q$ range from \SI{0.003}{} to \SI{0.4}{\angstrom^{-1}}, where $q = \frac{4 \pi \sin \theta}{\lambda}$ is the scattering wave vector and $2\theta$ is the scattering angle. All measurements were conducted in \SI{1}{} or \SI{2}{\milli\meter} quartz cells, depending on the D$_2$O fraction. As a reference, aqueous samples of PEG were analyzed by SANS. Two types of quartz cells were used depending on the samples. PEG solutions were introduced into classical cuvette cells, whereas membranes were inserted between two windows accurately spaced \SI{1}{\milli\m} apart. Measurements of the transmission, scattering of the empty cell, scattering of cadmium (used as a neutron absorber to estimate the ambient background of the SANS experiments), scattering of hydrogenated water, and the differential scattering cross-section of water were conducted independently. The raw data were processed using the GRASP software, which involves the subtraction of the parasitic contributions and the normalization with water to account for the detector heterogeneities, yielding corrected data in absolute units $(\mathrm{cm}^{-1}$). Subsequently, contributions from the solvent and incoherent scattering were also subtracted. All the obtained scattering curves and their corresponding fits are presented in the Supporting Information (SI Section 3).
It is known that the scattered intensity $I$ of objects at a scattering vector $q$ is given by\cite{cousin2015small}:
\begin{equation}\label{eq:I(q)}
    I(q) = \Phi \cdot \Delta \rho^2 \cdot V_{\textrm{obj}} \cdot P(q) \cdot S(q)
\end{equation}

\noindent where $ \Phi $ is the volume fraction of the object, $ V_{\mathrm{obj}} $ its volume, and  $\Delta \rho = \rho_{\mathrm{obj}} - \rho_{\mathrm{solvent}}$, the contrast term, where $ \rho_{\mathrm{obj}} $ and $ \rho_{\mathrm{solvent}} $ are the respective scattering length densities of the object and the solvent. $ P(q) $ is the form factor of the object, which contains information about the correlations between its internal elementary scatterers, while $ S(q) $ is the structure factor, which describes the correlations between the elementary scatterers from different objects. 

To achieve the best signal-to-noise ratio, the goal is to maximize the scattered intensity ($I$) by optimizing the contrast $ \Delta \rho $.  The scattering length density values of the compounds used in this study\cite{le2015conformation} are reported in the Supporting Information (SI Fig. S1) and show that D$_2$O exhibits a good contrast with PEG-h and is thus an appropriate solvent for the study of PEG-h solutions. However, in the case of the PEGDA/PEG-h membranes in D$_2$O, the PEG spacers of the PEGDA matrix and the added PEG chains display similar scattering length densities hence their contributions can not be distinguished. In such a kind of ternary systems, the PEGDA matrix or the PEG conformation may be selectively investigated by using the contrast matching method. To determine the structure of the PEGDA matrix the scattering signal of the added PEG chains need to be matched. To this end the free PEG-h chains are replaced by PEG-d in D$_2$O which enables to match the PEG-d contribution to the scattering and hence obtain a good contrast between the PEGDA matrix and D$_2$O.
To selectively obtain the conformation of the free PEG chains, a H$_2$O/D$_2$O mixture containing 21.7\% of D$_2$O is used as it enables to minimize the scattering signal of the PEGDA (See Section 3.1 of the SI). Hence incorporating PEG-d in a PEGDA matrix in such a H$_2$O/D$_2$O mixture enables to minimize the scattering of PEGDA and selectively obtain the scattering of the PEG-d chains. In the Supporting information (Fig. S1 and S2) are provided the values of scattering intensities of the PEGDA in a series of mixtures H$_2$O/D$_2$O of different compositions which enabled to choose the composition that enabled to minimize the PEGDA signal. 
In summary, the conformation and behavior of PEG-d has been studied using a solvent containing 21.7\% of D$_2$O, while the structure of the PEGDA membranes was obtained using pure D$_2$O. 
We note that the scattering intensities measured for PEG-h solutions in D$_2$O, those of PEG-d solutions in H$_2$O and those of PEG-d in the prepolymerization solutions are given in the Supporting Information (Fig. S3, S4 and S5 respectively). 

\subsection{Solid-state nuclear magnetic resonance (NMR) spectroscopy}

$^{1}H$ solid-state NMR experiments were performed using a 300 MHz Bruker Avance III HD NMR spectrometer and a 4 mm double-resonance Magic-Angle Spinning (MAS) probe. For all measurements, the sample temperature was regulated at $25^\circ$C. $^{1}H$ single-pulse experiments were carried out with and without MAS. $^{1}H$ transverse relaxation signals were recorded at a MAS spinning frequency of 2 kHz using the Hahn echo pulse sequence. The $90^\circ$($^{1}H$) pulse length was equal to $3.2 \textrm{s}$, and the recycle delay was set to 2 s, in agreement with the $^{1}H$ spin-lattice relaxation time values, $T_1$($^{1}H$). Two consecutive measurements of the $^{1}H$ transverse relaxation functions were systematically performed to ensure that sample rotation does not affect the NMR results.

\section{Experimental results}

A qualitative observation of the membranes shows that pure PEGDA membranes are transparent, while PEGDA/PEG-d membranes appear turbid. This indicates significant structural changes 
when PEG-d chains are present in the PEGDA matrix. In this study, the structure and dynamics of both PEGDA matrix and PEG chains are investigated using SANS and NMR experiments carried out on systems with a given PEGDA volume fraction and varying the PEG one. In Section 3.1, we use PEG-d chains in $\mathrm{D_2O}$ which enables us to focus on the structure and dynamics of the PEGDA matrix while in Section 3.2 we focus on the structure of PEG-d chains.  

\subsection{Structure and dynamics of the PEGDA matrix in the presence of added PEG-d chains}

 \subsubsection{SANS measurements}
 
To obtain quantitative information about the structure of the PEGDA network within the membrane, PEGDA membranes containing PEG-d chains prepared with D$_2$O as the solvent are examined for SANS measurements. In these conditions, the contribution from the PEGDA matrix to the scattered intensity is the only one to be considered. 

\begin{figure}[ht]
  
 \includegraphics[width=0.9\linewidth]{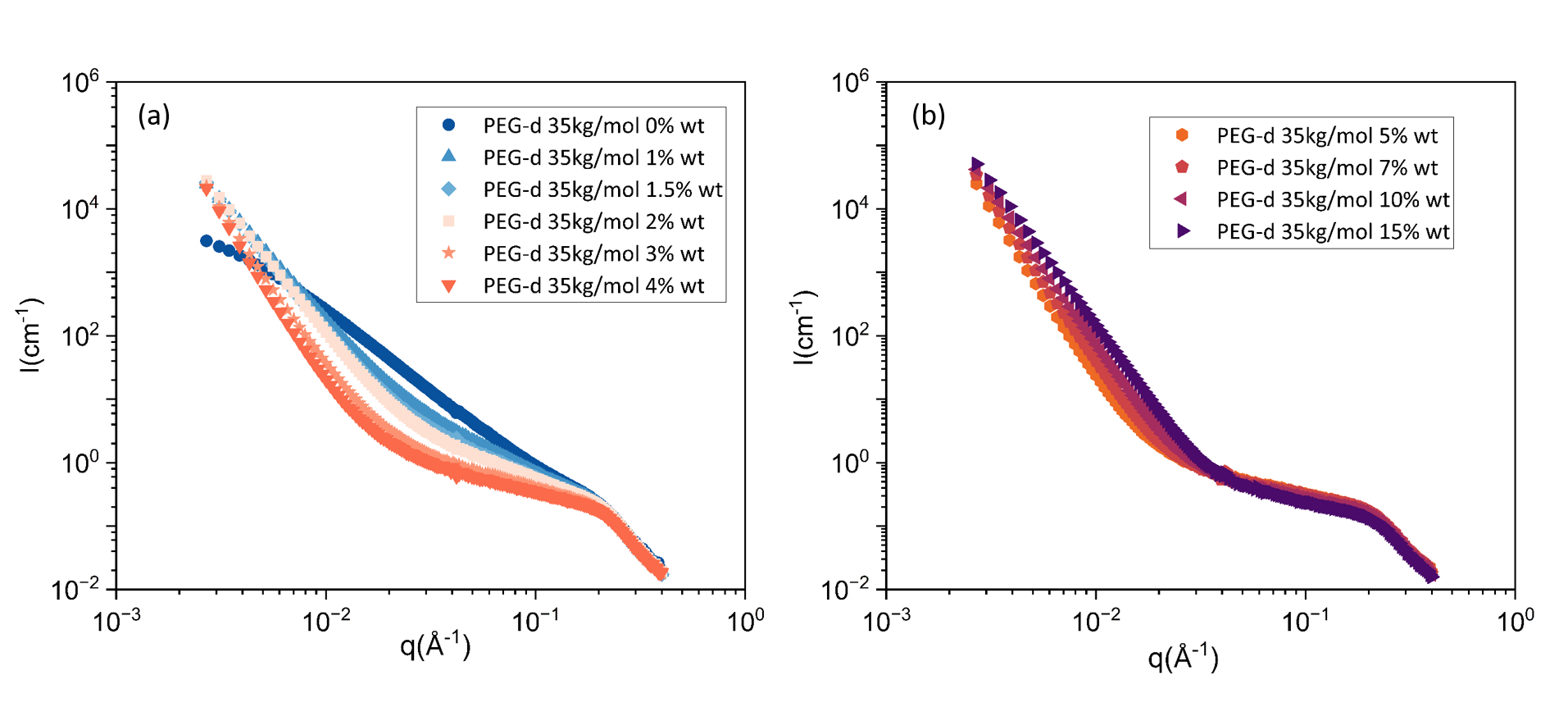}
   \caption{Scattering intensities as a function of wave vector $q$ for PEGDA/PEG-d samples in ($D_2\mathrm{O}$. a) PEG concentrations between 0 and 4 wt$\%$. b) PEG-d concentrations between 5 and 15 wt$\%$.}
   \label{fig:sans-PEGDA}
\end{figure}

The scattering profiles are presented in Figs. \ref{fig:sans-PEGDA}a and b for a PEGDA matrix containing added PEG-d chains of molar weight \SI{35}{\g\per\mol}, the concentrations of which standing below and above the overlap concentration of the PEG chains in solution ($C^*$) respectively. Additional scattering curves corresponding to samples containing PEG-d chains of molar weight \SI{3.5}{\g\per\mol} can be found in the Supporting Information (Fig. S6).

For a PEGDA membrane prepared without any added PEG-d chains at high \textit{q} values ($q > \SI{0.2}{\angstrom^{-1}}$), the I(q) curve exhibits a $q^{-3.8}$ power law for \textit{q} values below $q_{\text{c}} = \SI{0.2}{\angstrom^{-1}}$. For intermediate scattering vectors, the scattered intensity follows a $q^{-2.6}$ behavior and tends to reach a plateau for \textit{q} values below $\SI{0.01}{\a}$ . Such a scattering curve closely resembles the one reported by Molina \textit{et al}.\cite{DeMolina} for a similar PEGDA sample (i.e. a PEG spacer of length \SI{700}{\g\per\mol} and similar volume fraction of water of 80 wt\%). We summarize here their findings. They attributed the observed $q^{-3.8}$ behaviour at \textit{q} values above  $q_{\text{c}} = \SI{0.2}{\angstrom^{-1}}$ to the star-like structure of the PEG spacers of the PEGDA \cite{DeMolina}. The distance \textit{d}, with $d = \frac{2\pi}{q}<\SI{30}{\a}$, corresponds to the length of the PEG spacers in the PEGDA network. The $q^{-2.6}$ behaviour at intermediate \textit{q} values was attributed to the fractal dimension of the aggregated network of these stars. At the lowest \textit{q} values the curves level off to a plateau value. The latter observation was attributed by the authors to the presence of 200 nm cavities filled with water. According to these authors \cite{DeMolina} and others \cite{Ju2009}, the water voids are induced by a polymerization-induced phase separation obtained when the total volume fraction of water in the prepolimrization solution exceeds the equilibrium volume fraction that the PEGDA matrix can sustain. Indeed, the equilibrium volume fraction of the PEGDA/water matrix was found to be around 50/50 \cite{DeMolina}. Consequently, for the samples that contain a higher water volume fraction, the excess of water is found inside these cavities trapped in the PEGDA matrix. \\

When PEG-d chains are present during the formation of the PEGDA matrix, all the \textit{I(q)} curves overlap for \textit{q} values above \SI{0.2}{\angstrom^{-1}}, corresponding to short length scales ($d = \frac{2\pi}{q}<\SI{30}{\a}$). Thus, the presence of additional PEG-d chains in the PEGDA matrix does not influence its structure at the scale of the PEG spacers. In contrast, below \SI{0.2}{\angstrom^{-1}} the scattered intensity  depends on the PEG-d concentration, indicating that the PEG-d chains modify the PEGDA network at a scale larger than 30 angströms. As the PEG-d concentration increases below $C^*$  the exponent of the \textit{I(q)} curve gradually decreases from ${-2.6}$ to ${-4}$ and no leveling of the curves is observed contrary to the PEGDA samples with no PEG-d. At higher concentrations, the $q^{-4}$ behaviour observed for \textit{q} < \SI{0.01}{\angstrom^{-1}}, hence \textit{d} > 60 nm, indicates the formation of large heterogeneities possibly induced by a phase separation between PEGDA-rich and PEGDA-poor zones. In this regime, the scattering intensity can be written according to the Porod's law which writes as follows \cite{hammouda2004insight}:

\begin{equation}\label{eq:porod}
    {I(q)} = {x}_\mathrm{{rich}}^\mathrm{{PEGDA}} .\frac{2 \pi (\mathrm{\rho}_\mathrm{{rich}}^\mathrm{{PEGDA}} - \rho_\mathrm{{poor}}^\mathrm{{PEGDA}})^2}{{{q}^4}}\cdot \frac{S}{V}
\end{equation}

\noindent where $x_\mathrm{{rich}}^\mathrm{{PEGDA}}$ is the volume fraction of the PEGDA-rich phase, characterized by the scattering length density $\rho_\mathrm{{rich}}^\mathrm{{PEGDA}}$. $\rho_\mathrm{{poor}}^\mathrm{{PEGDA}}$ is the scattering length density of the PEGDA-poor phase and $S/V$ is the specific surface of the PEGDA heterogeneities. 
In the Discussion Section (Section 4), we use Eq. 2 to calculate $S/V$ values for the PEGDA heterogeneities as a function of the concentration of added PEG and relate these values to the permeability of the PEGDA/PEG membranes reported in our previous studies \cite{eddine2022large, eddine2024tuning}.

\subsubsection{Solid-state NMR}

The dynamics of the chain segments of the PEGDA membrane was investigated through  $^1$H NMR transverse relaxation, considering the PEGDA membranes in $\mathrm{D}_2 \mathrm{O}$ without PEG-d chains. Then, the response of the protons of the PEGDA spacers in the presence of PEG-d chains in $\mathrm{D}_2 \mathrm{O}$ was analysed. These conditions enabled to obtain the signal of the protons of the PEGDA network selectively. 

The $^1$H NMR spectrum of the PEGDA hydrogel without any added PEG chains was recorded under static
conditions. The contribution from the protons of the PEG spacers of PEGDA could hardly be detected (Figure 2). The slow MAS
spinning of the sample (2 kHz) lead to an efficient averaging of the $^1$H-$^1$H dipolar couplings $D_\mathrm{HH}$
and therefore, allowed the PEG spacer peak to be detected. This first result indicates that the protons of the
PEG spacers display $^1$H-$^1$H dipolar couplings as large as 2 kHz. In other words, 
repeat units in the PEGDA-based hydrogel display strongly-constrained, anisotropic
reorientational motions over the tens of microseconds time scale.

\begin{figure}[ht]
\centering
\includegraphics[width=0.6\linewidth]{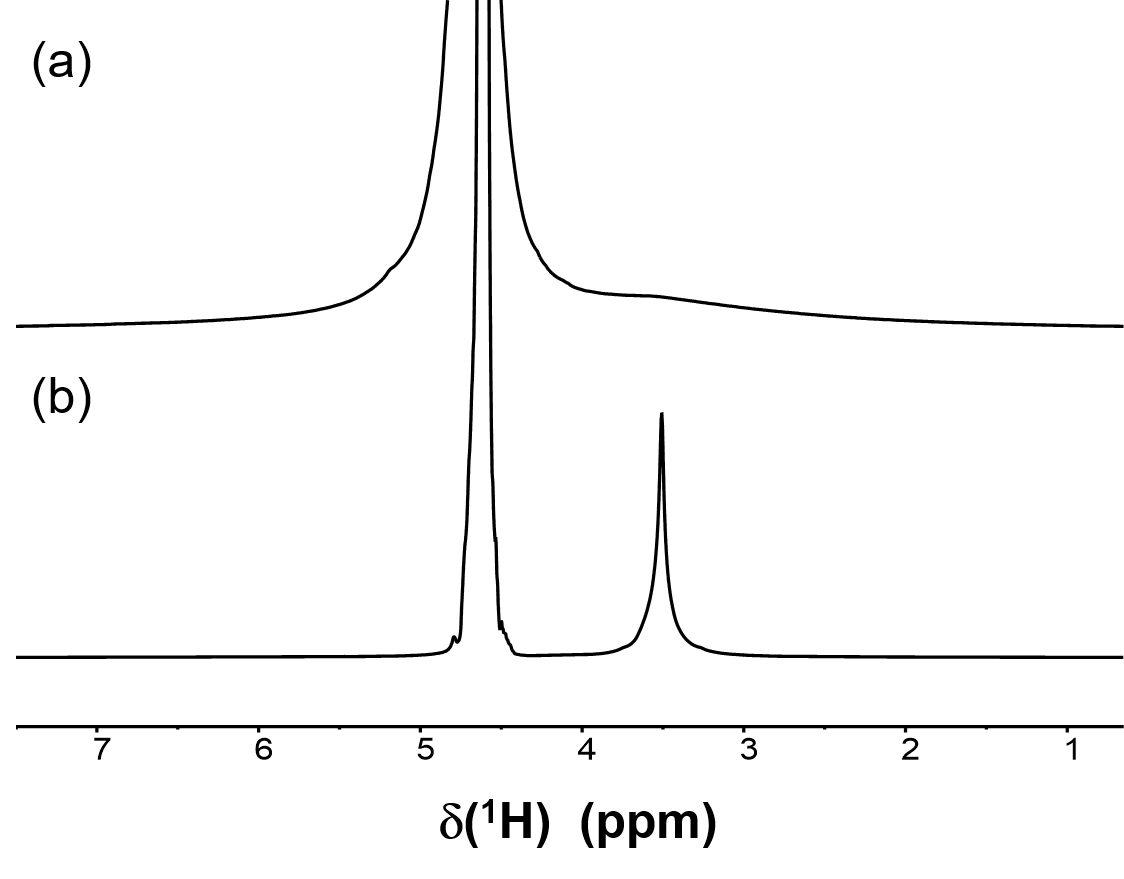}
\caption{$^1H$ NMR spectra of the PEGDA hydrogel at 25°C under static conditions (a) and under spinning at 2 kHz (b).}
\label{fig:RMN-statiqueversusspinning}
\end{figure}

The $^1$H transverse relaxation signal, $M(2\tau)/M_0$, of the PEG protons was determined at a MAS spinning 
frequency $\nu_r$=2 kHz (see Figure 3). At this stage, it is worth noting that the $^1$H MAS NMR
spectrum of the PEGDA hydrogel does not exhibit the contribution from vinyl protons, meaning
that all the PEG spacers are constrained by their extremities to the PEGDA backbone. In other
words, the fraction of dangling or uncrosslinked PEG spacers may be considered
as negligible. In this context, the function $M(2\tau)/M_0$ and, in particular, its deviation from a
monoexponential decay, may be assigned to the gradient of segmental mobility along the PEG
spacers: the motions from the repeat units at the PEG spacer extremities are more constrained
and as a result, the corresponding protons should display a faster relaxation. The units located
in the middle of the PEG spacers should be comparatively more mobile and the corresponding
protons are expected to display a slower $T_2$ relaxation. Such a gradient of segmental mobility for the protons along the PEG spacer my be described using a stretched exponential \cite{fengler2022situ}:

\begin{equation}\label{eq:nmr}
M(2\tau)/M_0 = \mathrm{exp}[-(2\tau/T_2)^{\beta}].
\end{equation}

The corresponding fit, obtained for $T_2$ = $15.6$ $ms$ and $\beta = 0.5$, is included in Figure 3.
The data obtained by $^{1}H$ Hahn echo experiments are consistent with the $^{1}H$ double-quantum build-up curve obtained at a MAS spinning frequency of 5 kHz through the BaBa-xy16 pulse sequence (Figure S7). In particular, the comparison between the reference signal,
$S_{Ref}$($t_{DQ}$), and the double-quantum signal, $S_{DQ}$($t_{DQ}$), indicates a negligible fraction of PEG repeat
units displaying isotropic reorientational motions over the tens of microseconds time scale.
Besides, the distribution of the $^{1}H$ dipolar coupling $D_{HH}$ leads to a mean value of 494 Hz,
consistent with the efficient line narrowing of the $^{1}H$ NMR peak from the PEG protons observed
at $\nu_r$ = 2 kHz.

In a second step, the $^{1}H$ transverse relaxation signal was recorded under the same
experimental conditions for PEGDA-based hydrogels prepared in the presence of a varying
fraction of deuterated PEG chains, ranging from 3 to 10 wt \% (Fig. 3). 
In these experiments, $M(2\tau)/M_0$
results from the contribution of the PEG repeat units forming the PEGDA network only. It may
be first remarked that for 3 wt \% of deuterated PEG chains, $M(2\tau)/M_0$ displays a significantly
faster relaxation than the one measured for the PEGDA-based hydrogel, prepared without any
added PEG chain. The fit of the experimental data using Eq. (1) leads to $T_2$ = $3.5$  ms  and $\beta$ = 0.38. The mean $\mathrm{T_2}$ relaxation time for the PEG repeat units of the PEGDA-based network
formed with 3 wt \% of free PEG-d chains is equal to 13 ms, to be compared with the value
determined for the neat PEGDA network, i.e. 63 ms. Such a decrease of the average $T_2$ value,
together with the reduction of the $\beta $ exponent, suggests that the segmental dynamics of the
PEG spacers from the PEGDA network gets more constrained and more heterogeneous
as the network is prepared with added deuterated PEG-d chains. This result may be induced by entanglements formed by the PEG-d chains with the PEGDA network
(number of repeat units per PEG-d chain, $N_\mathrm{{PEG-d}}$ = 729 against $N_\mathrm{{PEGDA}} = 13$ for
PEGDA). It should be noted that the $^{1}H$ transverse relaxation signals are discussed as governed
by $T_\mathrm{2}$ relaxation only (Eq. (3)). The anisotropic reorientational motions
lead to residual $^{1}H$ dipolar couplings, $D_\mathrm{{HH}}$, that also contribute to the decay of $M(2\tau)/M_0$. In a
first and rough approach, we did not attempt to discard the respective roles played by $T_2$ and
$D_\mathrm{{HH}}$ in the analysis of the Hahn echo measurements. 
Figure 3 also shows that above 3 wt\% of free PEG-d chains, the $^{1}H$ transverse
relaxation signal from the PEG protons involved in the PEGDA network remains unchanged,
within the experimental accuracy. 
\begin{figure}[ht]
\centering
\includegraphics[width=0.9\linewidth]{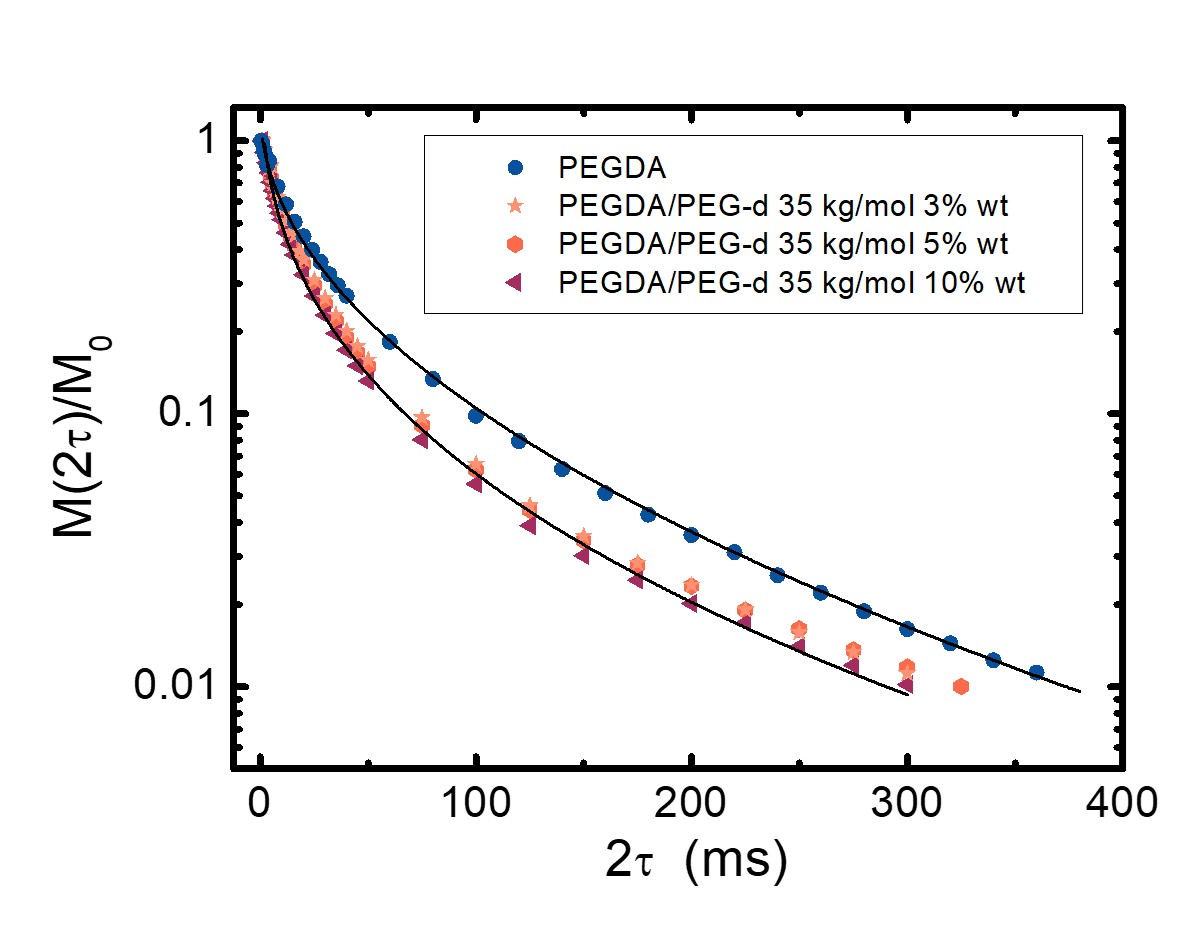}
\caption{$^{1}H$ transverse relaxation signal \textit{M}(\textit{2}$\tau$)/$M_0$ of the PEG protons of the PEGDA matrix with 0, 3, 5 and 10 wt\% of PEG-d chains. These data were recorded at 25 °C using the Hahn echo pulse sequence under MAS of the sample. All the echo amplitudes $\textit{M}(2\tau)$ were normalized by the one of the echo formed at the shortest relaxation delay used,$M_{0}$. The solid lines stand for a fit of the experimental data using a stretched exponential function}
\label{fig:RMN-deutere}
\end{figure}

\newpage

To summarize, the results presented above show that the structure of the PEGDA matrix at the nanometric scale is not affected by the presence of the PEG-d chains. However the dynamics of the PEG protons from the PEGDA matrix seems to be slowed down in the presence of the PEG-d chains suggesting possible entanglements between the PEG-d chains and the PEGDA network. At scales larger than hundreds of nanometers large heterogeneities in the PEGDA concentration are evidenced by SANS.

\subsection{Conformation and dynamics of the free PEG-d chains in the PEGDA membrane}

 \subsubsection{\textbf{SANS using PEG-d chains in a $\textbf{H}_\textbf{2}\textbf{O} / \textbf{D}_\textbf{2}\textbf{O}$} \textbf{ mixture}}

 The goal of this section is to determine the conformation of the added PEG chains within the PEGDA matrix. Under contrast-matching conditions, neutron scattering experiments selectively probed the PEG-d component within the PEGDA matrix. 
 
In Fig. 4 we present the scattering spectra obtained for PEG \SI{35}{\g\per\mol} at concentrations above 4 wt\%. Additional scattering curves corresponding to PEGDA matrix containing PEG-d chains of molar weight \SI{3.5}{\g\per\mol} can be found in the Supporting Information (Fig. S8). 

Below $C^{*}$ the signal obtained for PEG \SI{35}{\g\per\mol} is noisy (see Fig. S9 in the Supporting Information) and can be fitted 
with a $q^{-\alpha}$ slope with $\alpha \approx{1.15-1.5}$ meaning that the chains have a locally extended conformation. The $I(q)$ do not present a plateau at low $q$ which seems to indicate a slight aggregation of the chains. In comparaison the signal obtained for the PEG-d chains in solution present a $q^{-2}$ regime showing that they are in a theta solvent (Fig. S3, S4, S5, S10 and S11).

Above $C^{*}$ the scattering profiles shown in Fig. 4 were analyzed using a combined Porod and Lorentzian-like model, following the formalism proposed by Hamouda \textit{et al}. for polymer solutions \cite{hammouda2004insight}: 
\begin{equation} \label{Lorentz}
{I(q)}=\frac{A}{q^4}+\frac{B}{1+(q.\xi_{app})^m}.
\end{equation}
This model allows the extraction of an apparent correlation length $\xi_{app}$, associated with polymer concentration fluctuations, together with the Lorentzian exponent $m$ which characterizes the decay of the scattering intensity at intermediate and high scattering vectors. 

Within this framework, the exponent $m$ is related to an effective excluded-volume parameter $\mu$ through the relation $m = \frac{1}{\mathrm{\mu}}$, as discussed by Hammouda \textit{et al}. \cite{hammouda2004insight}. The parameter $\mu$ reflects the conformational statistics of polymer chains and provides insight into the effective solvent conditions experienced by the polymer at the length scales probed by SANS. For 4 wt$\%$, the fitted Lorentzian exponent is $m \approx 1.3$ corresponding to an effective excluded-volume parameter 0.77. This value is indicative of strongly swollen PEG chains and enhanced excluded-volume effects. In this regime, PEG-d chains are highly extended within the PEGDA matrix and experience limited screening of their interactions. The apparent correlation lengths extracted from the Lorentzian term $\xi_\mathrm{{app}}\approx 126$ $\text{\AA}$ are much higher than the correlation lengths $ \xi  \approx   8$ to 20 $\text{\AA}$ obtained for semi-dilute solutions of the same PEG-d chains in $\mathrm{H}_2\mathrm{O} $ at comparable concentrations (See Fig. S12 in Supporting Information).  
As the PEG concentration increases up to 15 wt\%, the exponent gradually increases to $m\approx 1.6$, corresponding to $\mu\approx 0.63$. This evolution suggests a progressive screening of excluded-volume interactions as the local PEG-d concentration increases. Concomitantly, the apparent correlation length decreases significantly, reaching $\xi_\mathrm{{app}}\approx 33$ $\text{\AA}$. This reduction of $\xi_\mathrm{{app}}$  reflects a decrease in the characteristic length scale of concentration fluctuations, consistent with a denser and more constrained polymer organization.
In this more concentrated regime, PEG-d chains increasingly interact with neighboring PEG-d chains and with the PEGDA network. As a result, chain conformations become more constrained and partially entangled and the local environment experienced by a given chain becomes increasingly rich in PEG-d. From the perspective of an individual chain, this situation corresponds to an effective solvent condition progressively approaching theta-like behavior\cite{hammouda2004insight}.
While both \textit{m} and $\xi_\mathrm{{app}}$  are model-dependent parameters, their systematic and correlated evolution with PEG-d concentration provides strong evidence for a continuous modification of polymer organization within the concentrated phase. This behavior is fully consistent with the scaling arguments and SANS analyses reported by Hammouda \textit{et al}. \cite{hammouda2004insight}, where changes in the Lorentzian-like parameters are associated with concentration-induced screening effects and modifications of polymer conformations rather than with a transition to poor solvent conditions. \\
The existence of a $q^{-4}$ regime at low \textit{q} values is due to the existence of large heterogeneities in the PEG-d concentration. The specific surface, $S/V$, of these heterogenities can be obtained with the Lorentz fit to our data using the following Porod Equation below :

\begin{equation}\label{eq:porod-peg}
    I(q) = x_\mathrm{{rich}}^\mathrm{{PEG-d}} \frac{2 \pi (\rho_\mathrm{{rich}}^\mathrm{{PEG-d}} - \rho_\mathrm{{poor}}^\mathrm{{PEG-d}})^2}{q^4}\cdot \frac{S}{V}
\end{equation}\\
\noindent where $x_\mathrm{{rich}}^\mathrm{{PEG-d}}$ is the volume fraction of the PEG-rich phase, characterized by the scattering length density $\rho_\mathrm{{rich}}^\mathrm{{PEG-d}}$, and $\rho_\mathrm{{poor}}^\mathrm{{PEG-d}}$ is the scattering length density of the PEG-poor phase.
In the Discussion (Section 4) we calculate the specific surfaces of the PEG-d concentration heterogeneities.
\\

\begin{figure}[ht]
\centering
\includegraphics[width=0.6\linewidth]{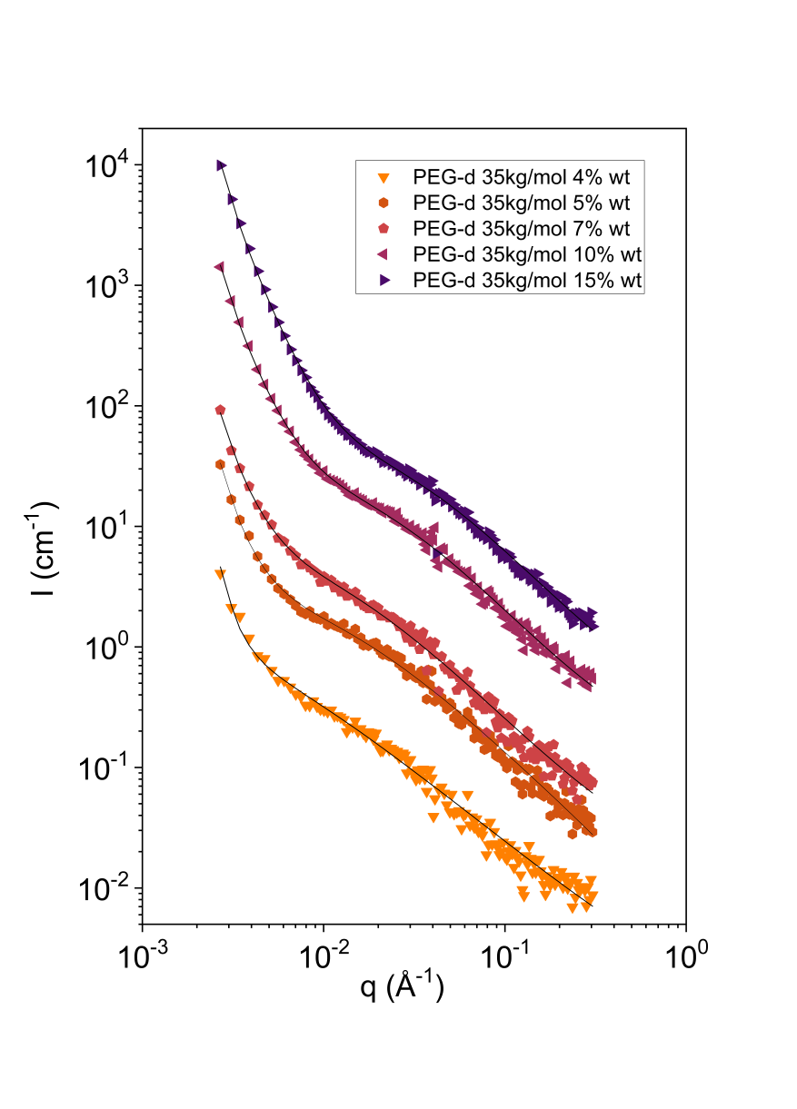}

\caption{Scattered intensities of the $\mathrm{PEG}_\mathrm{d} / \mathrm{PEGDA}$ samples in a $\mathrm{H}_2\mathrm{O}/\mathrm{D}_2\mathrm{O}$ mixture as a function of the wave-vector \textit{q}  for $ \mathrm{PEG}_\mathrm{d}$ concentrations from 4 wt\% to 15 wt\%- a multiplying factor is applied for each curve from bottom to top: x1; x5; x10; x50; x100}.
\label{fig:sans}
\end{figure}

\begin{table}
\centering

\begin{tabular}{| l | l | l | l |}
\hline
  wt$\%$& $\xi_{\mathrm{app}}$& $m$ & $\chi^2$ \\
\hline
4& 126 $\pm$ 27 & 1.3 $\pm$ 0.05 & 2.5 \\
\hline
5& 84 $\pm$ 1.6 & 1.5 $\pm$ 0.05 & 7 \\
\hline
7& 78 $\pm$ 1.5 & 1.5 $\pm$ 0.07 & 4 \\
\hline
10& 46 $\pm$ 0.5 & 1.6 $\pm$ 0.06 & 6.5 \\
\hline
15& 33 $\pm$ 1 & 1.6 $\pm$ 0.05 & 5 \\
\hline
\end{tabular}
\caption{Fitting parameters obtained when fitting the Porod-Lorentzian model to the scattering curves for $\mathrm{PEG}_\mathrm{d} / \mathrm{PEGDA}$ samples for $\mathrm{PEG}_\mathrm{d}$\ concentrations above 4 wt$\%$. $\xi_{\mathrm{app}} $ is the apparent correlation length and \textit{m} is the Lorentzian-like exponent related to the solvent quality.}
\end{table}

\subsubsection{Solid-state NMR}
\begin{figure}[ht]
\centering
\includegraphics[width=0.8\linewidth]{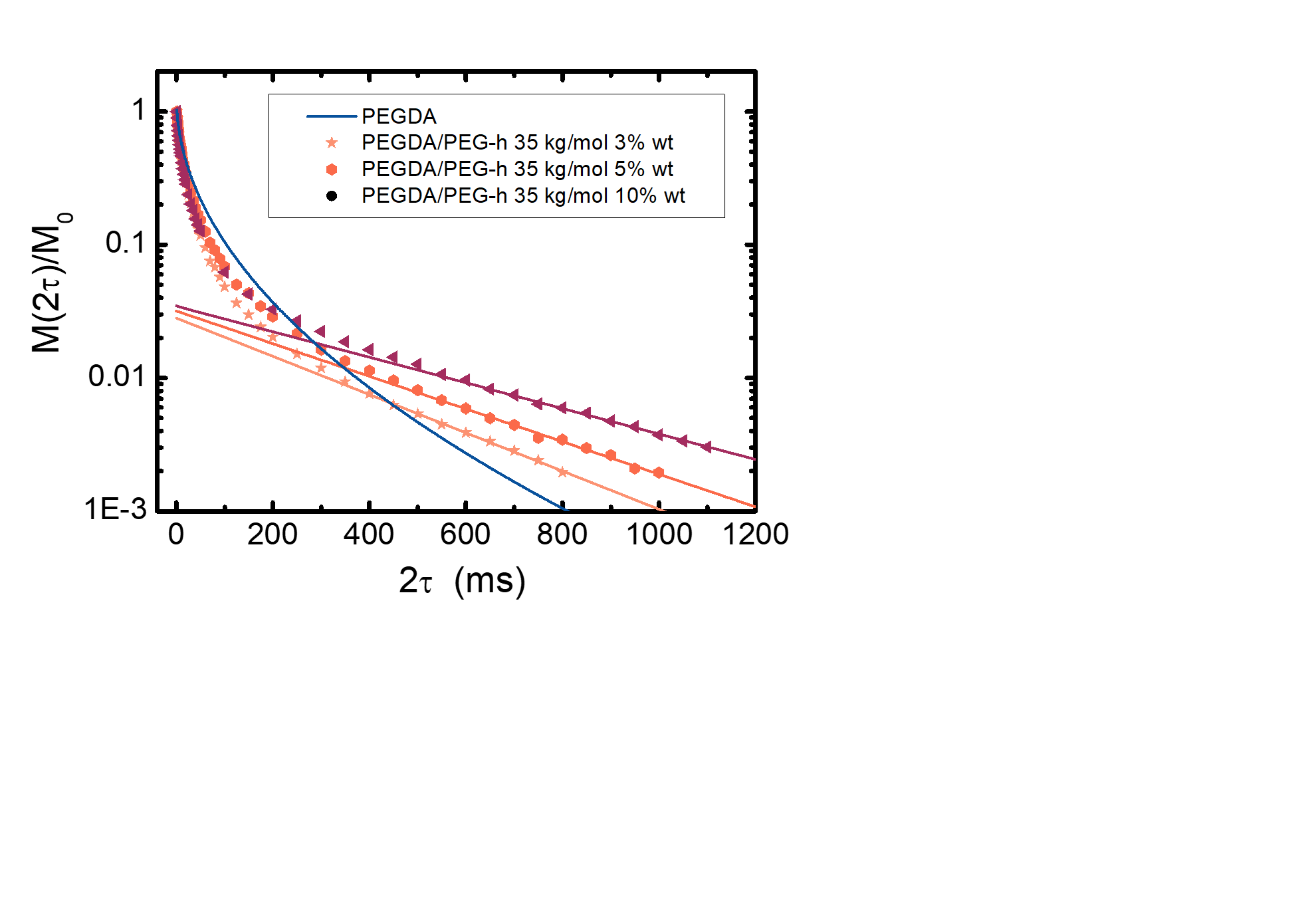}
\caption{1H transverse relaxation signal $M(2\tau)/M_{0}$ determined for the PEGDA/PEG membranes with 0, 3, 5 and 10 wt \% of PEG-h chains. These data were recorded at 25 °C using the Hahn echo pulse sequence under MAS. All the echo amplitudes $M(2\tau)$ were divided by the one of the echo obtained at the shortest relaxation delay used, $M_{0}$. The solid lines stand for a fit of the experimental data above 400 ms using a single-exponential function.}
\label{fig:RMN}
\end{figure}

The $^{1}H$ transverse relaxation signals were also recorded for the PEGDA-based hydrogels prepared with 3 wt$\%$, 5 wt$\%$ and 10 wt$\%$ of unlabeled PEG chains in $\mathrm{H}_2\mathrm{O}$, through the Hahn echo experiment carried out at a MAS spinning frequency of 2 kHz  (Figure \ref{fig:RMN}). The $^{1}H$ NMR peak at 3.5 ppm was again used so that in the present case, the relaxation functions correspond to the contributions from the PEG protons from both PEGDA network and added PEG chains. The amplitude of $M(2\tau)/\textit{M}_0$ was adjusted to take into account the additional amount of protons from the added PEG chains. From a qualitative point of view, the overall $^{1}H$ transverse relaxation measured with more than 3.33 \% of added PEG chains gets somehow faster than the one of the PEG spacers from the neat
PEGDA-based hydrogel. Again, this result is consistent with the fact that the added PEG chains interpenetrate with the PEGDA network: both PEG moieties from PEGDA and added PEG chains, simultaneously detected for these systems, entangle one with each other. These additional topological constraints result in an overall reduction of the segmental mobility and in an increased anisotropy of the reorientational motions of the PEG repeat units. Besides, Fig. 5 may suggest the occurrence of a slowly relaxing component which may be detected after 2$\tau$ = 400 ms. From a physical point of view, it could be assigned to a fraction of PEG repeat units from added chains that would display fast, isotropic reorientational motions over the tens of microseconds time scale. Such a situation could occur, for instance, for chain segments located in the pores of the PEGDA-based hydrogels. However, it should be pointed out that the corresponding fraction is rather weak, typically below 6 mol \% as the total concentration of PEG chains is raised from 3.33 wt \% to 10 wt \%. In all cases, most of the PEG repeat units from the added PEG chains seem to be entangled and constrained within the network.

\newpage

\section{Summary of the results and discussion}

\textbf{Structure at the scale of the PEG spacers and PEG-d chains}\\
Using SANS, we investigated the structure of the PEGDA network (Section 3.1.1) and found that the conformation of the PEG spacers at the nanometric scale is not modified in the presence of the PEG-d chains. However, the dynamics of the PEG spacers of the PEGDA matrix, probed by NMR (Section 3.1.2) is more constrained in the presence of free PEG-d chains. This first set of results lead us to hypothesize that the free PEG-d chains entangle with the PEGDA matrix thus reducing their mobility. Consistently, the NMR results presented in Section 3.2.2 showed that only a very small fraction of the protons of both free PEG chains and PEG spacers have a high mobility. From these results, we hypothesized that most of the PEG-d chains are entrapped within the PEGDA network. The extended conformation of the PEG-d chains observed with SANS (Section 3.2.1) suggests that the incorporation of the PEG-d within the PEGDA matrix leads to a more constrained conformation of both the PEG monomer units of the PEGDA network and of the added PEG chains.

\textbf{Structure at the scale larger than 60 nm - heterogeneities in the PEGDA and PEG concentrations}\\ 
The scattering spectra of the PEGDA matrix and the one of the PEG-d chains present a Porod regime at low q, scaling as $I \propto{q^{-4}}$, caused by the existence of large heterogenities in both the PEG-d and PEGDA concentrations. 
From Fig. 1 and Fig. 4 we determine the prefactors of this $I \propto{q^{-4}}$ portion of curve for PEGDA and PEG-d respectively, which enables us to calculate the specific surface $S/V$ of these heterogeneities using Eq. \ref{eq:porod} and Eq. \ref{eq:porod-peg} for PEGDA and PEG-d. To do so, we need to estimate $x_\mathrm{{rich}}^\mathrm{{PEG-d}} $ and $x_\mathrm{{rich}}^\mathrm{{PEGDA}}$, the total volume fractions of heterogenities rich in PEG-d and in PEGDA. We make the assumption of a coexistence between a phase rich in both PEG and PEGDA that coexists with a phase which only contains water, consistently with the previous results. Hence, we define the total volume fraction, $x_\mathrm{{rich}}$, of this phase rich in both PEG-d and PEGDA as follows : 

\begin{equation}\label{egalité}
 \\x_\mathrm{{rich}} = x_\mathrm{{rich}}^\mathrm{{PEGDA}} = x_\mathrm{{rich}}^\mathrm{{PEG-d}}.
\end{equation}\

The scattering length densites $\rho_{\mathrm{rich}}^\mathrm{PEGDA}$, $\rho_{poor}^{PEGDA}$,$\rho_{\mathrm{rich}}^\mathrm{PEG-d}$ and $\rho_{\mathrm{poor}}^\mathrm{PEG-d}$ of the phases rich and poor in PEGDA and those rich and poor in PEG-d are related to the volume fraction of each component (PEG-d, PEGDA and water) in each of the two phases according to: 
$$\rho_{\mathrm{poor}}^\mathrm{PEGDA} = \Phi_{\mathrm{poor}}^\mathrm{PEGDA}\cdot \rho_{\textrm{PEGDA}} + (1-\Phi_{\mathrm{poor}}^\mathrm{PEGDA}) \cdot \rho_{\text{D}_2\text{O}}$$

$$\rho_\mathrm{{rich}}^\mathrm{{PEGDA}} = \Phi_\mathrm{{rich}}^\mathrm{{PEGDA}}\cdot \rho_{\text{PEGDA}} + (1-\Phi_\mathrm{{rich}}^\mathrm{{PEGDA}}) \cdot \rho_{\text{D}_2\text{O}}$$

$$\rho_\mathrm{{rich}}^\mathrm{{PEG-d}} = \Phi_\mathrm{{rich}}^\mathrm{{PEG-d}}\cdot \rho_{\text{PEG-d}} + (1-\Phi_\mathrm{{rich}}^\mathrm{{PEG-d}}) \cdot \rho_{\text{D}_2\text{O}/\text{H}_2\text{O}}$$

$$\rho_\mathrm{{poor}}^\mathrm{{PEG-d}} = \Phi_\mathrm{{poor}}^\mathrm{{PEG-d}}\cdot \rho_{\text{PEG-d}} + (1-\Phi_\mathrm{{poor}}^\mathrm{{PEG-d}}) \cdot \rho_{\text{D}_2\text{O}/\text{H}_2\text{O}}$$

\noindent where $\rho_{\text{PEGDA}}$ and $\rho_{\text{PEG-d}}$ are the scattering length density of PEGDA and PEG-d, $\Phi_\mathrm{{rich}}^\mathrm{{PEGDA}}$ and $\Phi_\mathrm{{poor}}^\mathrm{{PEGDA}}$ stand for the volume fractions of PEGDA in the PEGDA-rich and -poor phases respectively and $\Phi_\mathrm{{rich}}^\mathrm{{PEG-d}}$ and $\Phi_\mathrm{{poor}}^\mathrm{{PEG-d}}$ stand for the volume fractions of PEG in the PEG-rich and PEG-poor phases. Moreover we note that $\rho_{\text{PEGDA}}=\rho_{\mathrm{{D}_2\mathrm{O}/\mathrm{H}_2\mathrm{O}}}$ and $\rho_{\text{PEG-d}} = \rho_{\text{D}_2\text{O}}$.

Moreover the total volume fractions in PEGDA, PEG-d and water, $\Phi_\mathrm{{TOT}}^\mathrm{{PEG-d}}$, $\Phi_\mathrm{{TOT}}^\mathrm{{water}}$, $\Phi_\mathrm{{TOT}}^\mathrm{{PEGDA}}$, can be defined as follows :
\begin{equation}\label{phitotpegda}
    \Phi_\mathrm{{TOT}}^\mathrm{{PEGDA}} = x_\mathrm{{rich}}\cdot\Phi_\mathrm{{rich}}^\mathrm{{PEGDA}}
\end{equation}
\begin{equation}\label{phitotpegd}
\Phi_\mathrm{{TOT}}^\mathrm{{PEG-d}}=x_\mathrm{{rich}} \cdot \Phi_\mathrm{{rich}}^\mathrm{{PEG-d}}
\end{equation}
\begin{equation}\label{phitotpegd}
\Phi_\mathrm{{TOT}}^\mathrm{{water}}=x_\mathrm{{rich}}\cdot \Phi_\mathrm{{rich}}^\mathrm{{water}} + (1-x_\mathrm{{rich}}).
\end{equation}

Hence we obtain :

\begin{equation}\label{xrich}
   x_\mathrm{{rich}}= \frac{\Phi^\mathrm{{water}}_\mathrm{{TOT}}-1}{\Phi^\mathrm{{water}}_\mathrm{{rich}}-1}
\end{equation}
\begin{equation}\label{phirichpeg}
    \Phi_\mathrm{{rich}}^\mathrm{{PEG-d}}=\frac{\Phi_\mathrm{{TOT}}^\mathrm{{PEG-d}}}{x_\mathrm{{rich}}}
\end{equation}
\begin{equation}\label{phirichpegda}
    \Phi_\mathrm{{rich}}^\mathrm{{PEGDA}}=\frac{\Phi_\mathrm{{TOT}}^\mathrm{{PEGDA}}}{x_\mathrm{{rich}}}.
\end{equation}

\begin{figure}
\centering
  \includegraphics[width=0.6\linewidth]{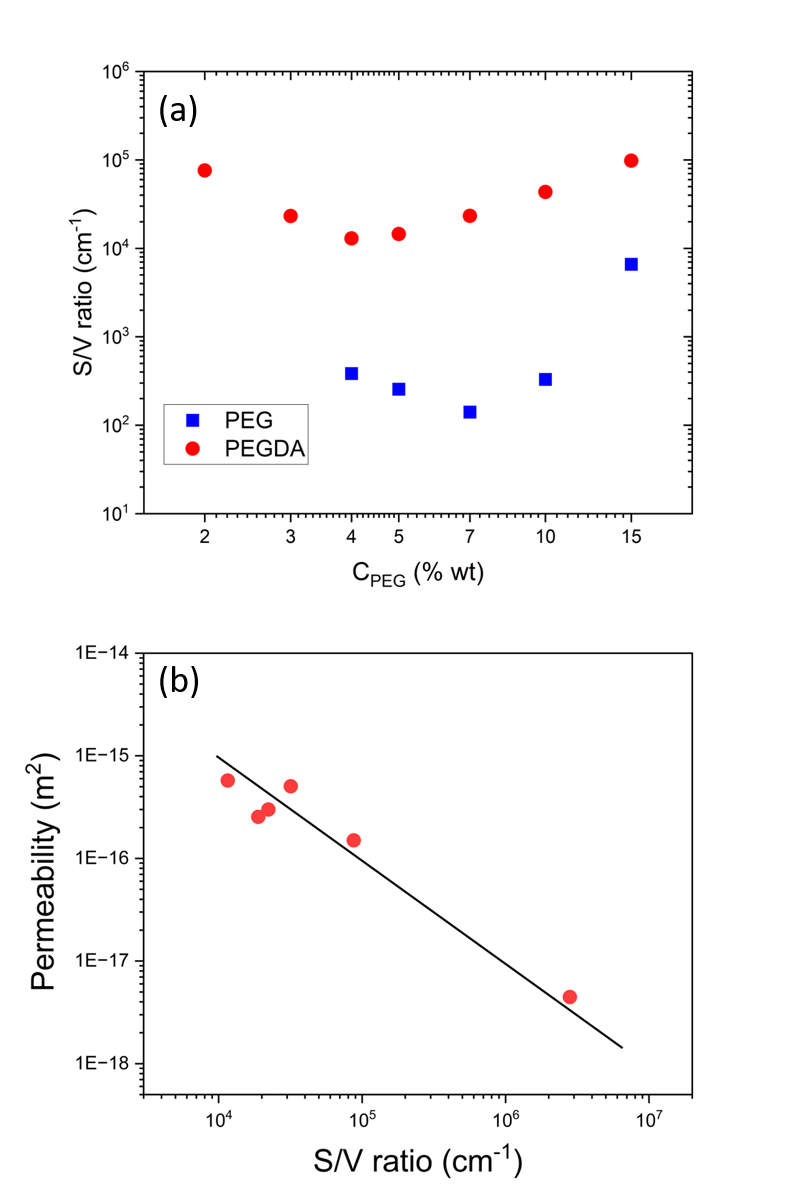}
 
\caption{a. Specific surface, $S/V$, of PEGDA matrix and of the PEG-d chains as a function of the PEG concentration; b. Variation of the permeability as a function of the $S/V$ measured for the PEGDA matrix, data taken from \cite{eddine2024tuning}}
\label{fig:ssurv}
\end{figure}

From the above equations we calculated the specific surfaces of both the PEGDA and PEG-d heterogeneities (Fig. \ref{fig:ssurv}a).  The $S/V$ calculated for the PEGDA presents a minimum for $C_\mathrm{{PEG}}=  C^*$. Moreover, the $S/V$ obtained for PEG-d, calculated for 4 wt\%, is two orders of magnitude lower than the one of the PEGDA at 4 wt\%. However, as the PEG-d concentration increases $S/V$ increases by orders of magnitude for the PEG-d chains and becomes closer to the values measured for PEGDA. This result supports the hypothesis according to which the PEG-d chains are incorporated continuously in the PEGDA matrix as the PEG concentration increases.

\textbf{Link between $S/V$ and permeability}

In our previous articles \cite{eddine2022large, eddine2023sieving, eddine2024tuning}, we have shown that the permeability of our PEGDA/PEG hydrogels presents an optimum with the PEG concentration, obtained at the overlap concentration of the chains in solution, \textit{C*}. In the present study we showed that the $S/V$ ratio, which is related to the hydrogel heterogeneities, presents a minimum with the PEG-d concentration at C*. It is therefore tempting determine whether the permeability variations observed previously are related to structural variations of the hydrogel membranes. Indeed in a porous medium, the permeability is controlled by the pore geometry due to the friction between the liquid and the walls leading to a decrease of the permeability with decreasing pore size and increasing tortuosity. We thus plot the permeability $K$, reported in \cite{eddine2024tuning} for PEG \SI{35}{\g\per\mol}, as a function of the $S/V$ ratio of the PEGDA matrix in Fig. \ref{fig:ssurv}. We obtain a master curve showing that the permeability scales as $K \propto {(S/V)^{-1}} $ both below and above C*. This result suggests that in our permeation experiments water probably permeates between the walls of the PEGDA/PEG-d heterogeneities and that the shape of these heterogeneities controls the permeability of the hydrogels. Taking a simple geometry, ie spherical beads of radius \textit{R}, $S/V \propto{3/R}$. From the measured values $S/V$$\approx  10^5$ $ \mathrm{cm}^{-1} - 10^4  $ $\mathrm{cm}^{-1}$  we estimate the channel sizes $R \approx 30 - 300 \mathrm{nm}$. However, the permeability of such a medium should scale as K $\propto{R^2} \propto{(S/V)^{-2}} $ \textit{i.e.} a power law with a higher exponent than the one that we observe. 
 This suggests flattened or facetted PEGDA / PEG domains in contact with a network of aqueous thin films. Indeed, in the extreme case of flat water films of radius R and height h, one has $S/V\propto{1/h}$ and $K\propto{R.h}\propto{R.V/S}$, which indicates and exponent -1 on S/V, that is consistent with the value measured experimentally from Fig. 6.

\textbf{Structure of the PEGDA/PEG-d hydrogels}\\
Our present results indicate a possible entanglement of the PEG-d chains with the PEGDA matrix and a very small fraction of protons with high mobility. These results are in favor of an intimate incorporation of PEG chains in the PEGDA matrix and thus seem in contradiction with our previous hypothesis according to which individual PEG chains induce in their vicinity defects in the cross-linking density of the PEGDA network \cite{eddine2022large}. We thus propose the following picture for the hydrogels (Fig. 7) : a coexistence of areas, of flattened shape, rich in PEGDA and PEG, and a network of aqueous flattened thin films. 

While in PEGDA pure samples without PEG, previous studies using SANS and cryoSEM have reported the presence of 200 nm water cavities in the continuous PEGDA matrix, we suggest that the presence of PEG induces a modification of the structure where the water cavities form a percolated network of water channels between the PEGDA/PEG-d rich areas. 

\begin{figure}[ht]
\centering
\includegraphics[width=0.8\linewidth]{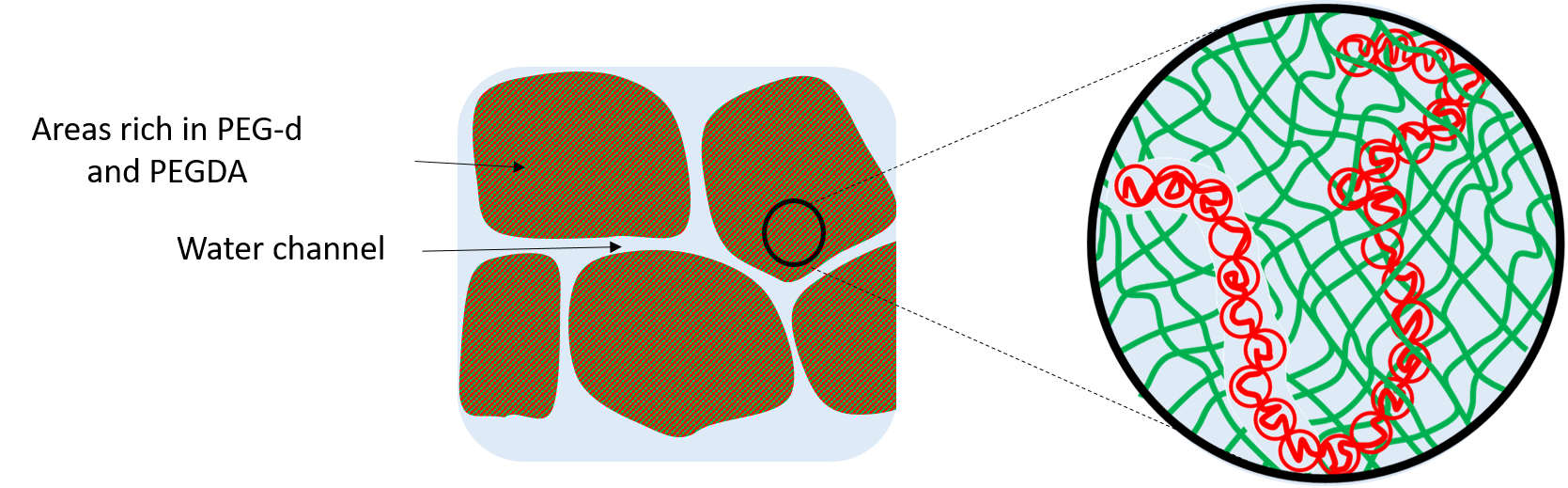}
\caption{Proposed schematic drawing of the PEGDA/PEG hydrogels with disconnected PEGDA/PEG-d rich domains coexisting with conitnuous flat cylindrical-like water channels.}
\label{fig:structure}
\end{figure}
 
These large heterogenities in the network composed of PEGDA and PEG-d can be viewed as a Polymerization Induced Phase Separation (PIPS) induced by the fact that the water volume fraction in the polymerization solution exceeds the equilibrium composition that the polymerized PEGDA/PEG matrix can sustain \cite{DeMolina}. The water in excess is therefore expelled from the PEGDA network and trapped into a continuous network of aqueous thin films. 

\section{Conclusion} 

Using SANS and NMR, we have investigated the structure of PEGDA/PEG hydrogels as a function of the PEG concentration below and above the PEG overlap concentration $C^*$ of PEG chains in solution. We obtained a set of data which shows that the hydrogels present heterogenities in the PEGDA/PEG concentrations. The PEG chains seem to be located mainly within the PEGDA matrix where they have an extended  conformation and a constrained dynamics suggesting that the PEG chains entangle with the PEGDA network. The specific surface of the PEGDA heterogeneities seems to control the permeability of water. The dependance of the permeability $K\propto{S/V}^{-1}$ suggests a materials composed of facetted PEGDA heterogeneities separated by a network of aqueous flattened films through which water can permeate.

\section{Acknowledgements}
C. M and B. B acknowledge Carnot IPGG Microfluidique (grant number C21-17-2021-036) for financial support as well as Doctoral School ED397 (Physique et Chimie des Matériaux) of Sorbonne University for the fellowships of S. de Chateauneuf Randon and M. Alaa Eddine. C. L. thanks the NMR facility of Sorbonne Université and the Région Ile-de-France for its contribution to the acquisition of the solid-state NMR spectrometer, in the framework of DIM Nano-K. The authors acknowledge financial support from the European Union through the European
Research Council under EMetBrown (ERC-CoG-101039103) grant. They also
acknowledge financial support from the Interdisciplinary and
Exploratory Research program under MISTIC grant at University of Bordeaux,
France. Besides, they also acknowledge the support from the Matter and Light Science Department at University of Bordeaux, and the Réseaux de Recherche
Impulsion “Frontiers of Life,” which received financial support from the French
government in the framework of the University of Bordeaux’s France 2030
program. Finally, they thank the Soft Matter Collaborative Research Unit, Frontier
Research Center for Advanced Material and Life Science, Faculty of Advanced
Life Science at Hokkaido University, Sapporo, Japan, and the CNRS International
Research Network between France and India on “Hydrodynamics at small scales:
From soft matter to bioengineering.”

\newpage

\bibliographystyle{unsrt}
\bibliography{biblio_art_structure}

\end{document}